\documentclass[a4paper,fleqn]{cas-dc}

\usepackage[numbers]{natbib}

\usepackage{float}
\usepackage{url}
\usepackage{placeins}
\usepackage{xcolor}
\usepackage[normalem]{ulem}
\usepackage{amssymb}
\usepackage{siunitx}
\usepackage{comment}

\begin{document}
\let\WriteBookmarks\relax
\def\floatpagepagefraction{1}
\def\textpagefraction{.001}

\shorttitle{Low-background and long-term decay-rate measurements at JURLab}
\shortauthors{F. Sprok et~al.}

\title[mode = title]{Preparation for Low-radiation Background and Long-term
  Decay-rate Measurements at the Jánossy Underground Research Laboratory}

\author[1]{Franciska Sprok}[orcid=0009-0001-7740-0091]

\author[1,2]{Edit Fenyvesi}[orcid=0000-0003-2777-3719]

\author[2,3]{Gábor Gyula Kiss}[orcid=0000-0002-6872-916X]

\author[1]{Dénes Molnár}[]

\author[1]{Péter Lévai}[orcid=0009-0006-9345-9620]

\author[1]{Gergely Gábor Barnaföldi}[orcid=0000-0001-9223-6480]
\cormark[1]
\ead{barnafoldi.gergely@wigner.hun-ren.hu}

\affiliation[1]{organization={HUN-REN Wigner Research Centre for Physics},
                addressline={Konkoly-Thege Miklós Road 29-33},
                postcode={1121},
                city={Budapest},
                country={Hungary}}

\affiliation[2]{organization={HUN-REN Institute for Nuclear Research},
                addressline={Bem Square 18/C},
                postcode={4026},
                city={Debrecen},
                country={Hungary}}

\affiliation[3]{organization={HUN-REN Office for Supported Research Groups},
                addressline={Alkotmány Street 29},
                postcode={1052},
                city={Budapest},
                country={Hungary}}

\cortext[cor1]{Corresponding author}

\begin{abstract}
We report on the preparation and characterization of a low-background
$\gamma$-counting facility at the Jánossy Underground Research Laboratory
(JURLab) of the HUN-REN Wigner Research Centre for Physics, located at a depth
of 75~m water equivalent. The $\gamma$-ray background was measured with a
high-purity germanium detector in several shielding configurations and compared
with published spectra of other underground laboratories. The present shielding
of 5--10~cm lead with a 2~mm copper liner suppresses the background by about two
orders of magnitude. The long-term stability of the setup was investigated with
a \textsuperscript{137}Cs source and \textsuperscript{40}K over the period from
July~2022 to April~2024 with continuous environmental monitoring. The
decay-corrected \textsuperscript{137}Cs/\textsuperscript{40}K photopeak-area
ratio shows no residual trend within uncertainties, demonstrating suitability
for precision long-duration measurements. With the present setup at JURLab
minimum detectable activities are derived for radionuclides relevant to nuclear
astrophysics and nuclear technology. These results establish JURLab as a
suitable site for decay-rate studies and for low-activity measurements of
activation samples produced at nearby research facilities.
\end{abstract}


\maketitle

\section{Introduction}
\label{sec:intro}

Radioactive decay is one of the most fundamental processes in nuclear physics. In the standard description, the number of undecayed nuclei decreases exponentially with time as $\sim \exp(-\lambda t)$, where the decay constant $\lambda$ characterizes the nucleus under consideration. Precision measurements of decay rates therefore require 
sensitive testing of detector stability and environmental control. 
The question of whether radioactive decay rates are strictly constant under laboratory conditions has been revisited repeatedly over the last decades. Several long-term measurements have reported apparent time-dependent variations in measured count rates or in ratios of decay rates~\cite{ALBURGER1986168,Ellis_1990,Falkenberg2001,Parkhomov,Sturrock2011,JENKINS2009407,mcduffie2020}. However, such claims remain controversial, and in many cases it is difficult to disentangle possible physical effects from instrumental drift, environmental influences, or analysis choices. At the same time, several dedicated high-stability experiments have found no evidence for periodic variations at the level suggested by earlier reports~\cite{BELLOTTI2012114,BELLOTTI2013116}. This situation motivates further long-duration measurements performed under well-controlled conditions with careful monitoring of systematic effects.

It is well established that, in specific circumstances, decay properties can depend on the atomic or ionic environment. For example, electron-capture decays may exhibit host-material effects, and highly ionized nuclei can show dramatically modified decay properties compared to neutral atoms~\cite{Emry1972}. These examples demonstrate that the stability of a measured decay rate cannot be assumed under all experimental conditions and highlight the importance of distinguishing genuine physical effects from measurement artifacts. In the present work, however, our focus is not on these known environment-dependent modifications themselves, but on establishing an experimental setup suitable for testing possible time-dependent variations in a controlled underground laboratory environment.

A key challenge in long-term decay-rate studies is controlling systematic effects over timescales extending beyond several years. Small changes in gain, dead-time, background conditions, source positioning, temperature, humidity, radon concentration, or analysis window definition can all produce apparent variations in measured count rates. For this reason, a competitive decay-rate experiment requires not only low-statistical uncertainty, but also a stable detector response, reliable calibration, and continuous monitoring of environmental parameters. Underground operation is advantageous in this context because it reduces cosmic-ray-induced backgrounds and typically provides a more stable environment than surface laboratories. Several dedicated measurements performed at the Gran Sasso National Laboratory (LNGS) provide important experimental context for such studies. In particular, long-term measurements of \textsuperscript{137}Cs found no evidence for time-dependent modulation, excluding annual oscillation amplitudes larger than $8.5\times10^{-5}$ at 95\% confidence level~\cite{BELLOTTI2012114}. Additional studies of \textsuperscript{40}K, \textsuperscript{137}Cs, and \textsuperscript{232}Th in temporal coincidence with strong solar flares also found no significant deviations from standard expectations at the level of a few $10^{-4}$~\cite{BELLOTTI2013116}. These results underline both the difficulty and the importance of controlling systematics in long-term precision measurements.

Furthermore, there are several applications where low activities have to be determined precisely. For example, in nuclear astrophysics, the reaction cross-sections at energies corresponding to stellar temperatures are typically very low, often not exceeding a few $\mu$barns. For many relevant reactions, the experiments are performed using the activation procedure~\cite{Gyurky2019}. In this experimental technique, the nuclear reaction and the activity measurement are separated in time (and often) in space. A low-background laboratory provides an ideal location for activity measurements when studying reactions leading to final long-half-life nuclei. In a pilot project, we successfully investigated the cross section of the $^{100}$Mo($\alpha$,n)$^{103}$Ru reaction~\cite{Szegedi_2021idx}, with part of the activity measurements performed in the Jánossy Underground Research Laboratory (JURLab). Furthermore, precise cross sections are also required for applications in nuclear technology~\cite{Cano_Ott_2026}. For example the neutron radiation of the ($\alpha$,n) reactions is a serious concern for rare event search experiments~\cite{DarkSide, CRESST} that require very low levels of radioactive background. If the detector material is contaminated with naturally occurring $\alpha$-emitters such as $^{235}$U, $^{238}$U or $^{232}$Th, their decay is followed by the emission of several relatively energetic alpha particles which produce neutrons through ($\alpha$,n) reactions on light nuclei. For some relevant reactions (e.g. $^{19}$F($\alpha$,n)$^{22}$Na), the activation procedure may be a possible method for cross section determination if the activity measurement is performed in a laboratory with low $\gamma$-background.

\begin{figure*}[h!]
  \centering
  \includegraphics[width=.80\textwidth]{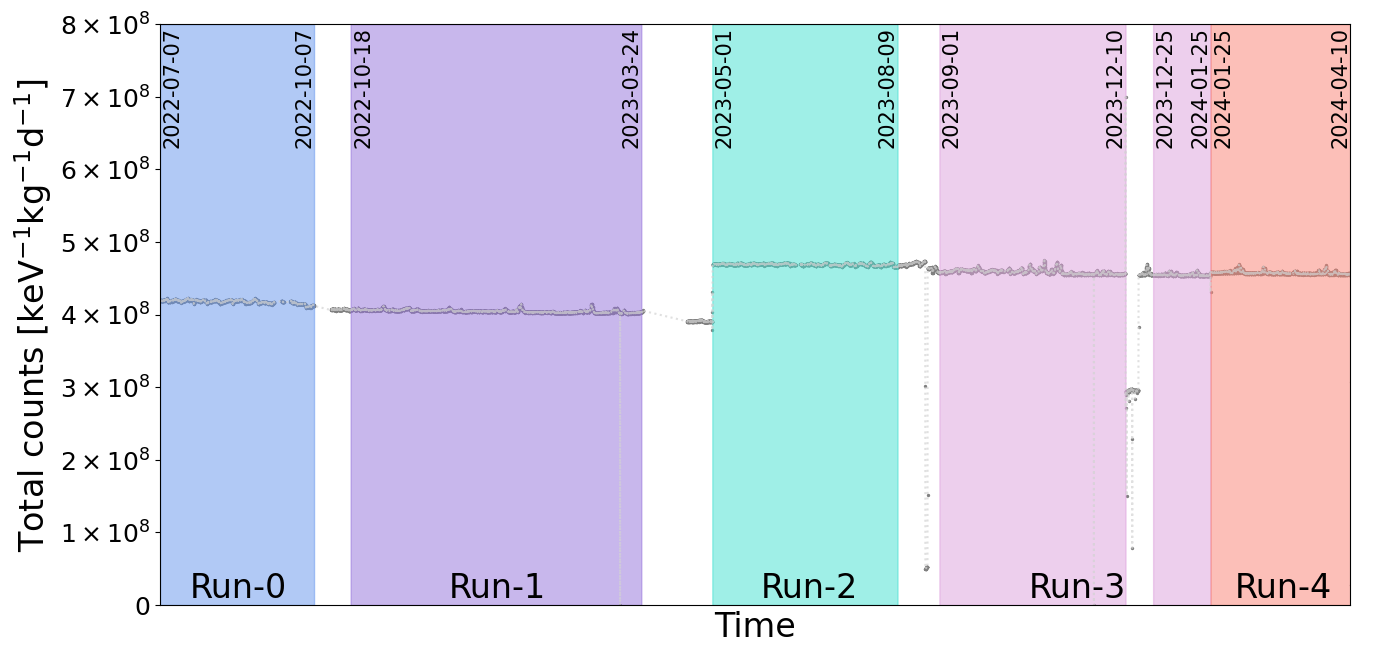}
  \caption{Timeline of the experiment showing total hourly counts (gray points) in the 40 keV - 2700 keV range recorded during every data-taking period. The duration of each run's measurement are indicated by colored regions.}
  \label{timeline}
\end{figure*}
In order to examine the current measurement conditions and determine the extent of the laboratory background, we initiated a long-term decay-rate measurement program at the JURLab of the HUN-REN Wigner Research Centre for Physics (HUN-REN Wigner RCP). The experiment uses a \textsuperscript{137}Cs source and a high-purity germanium detector installed in a low-background underground environment. The choice of \textsuperscript{137}Cs is motivated by its simple decay scheme, long half-life ($T_{1/2} = 30.018(22)$ years~\cite{Mougeot_2025}), and strong isolated $\gamma$ line at E$_{\gamma}$ = 661.7~keV, which make it well suited for stable long-duration counting measurements.
The purpose of the present paper is to describe the experimental setup and to report its initial characterization. In particular, we present the laboratory conditions, detector configuration, environmental monitoring, background measurements, and first studies of detector stability based on data collected between July~2022 and April~2024. The emphasis of this work is methodological: we aim to assess the stability of the detector response and to identify the dominant systematic effects that must be controlled in future high-sensitivity analyses. 
We also present the JURLab suitability for measuring isotopes of long-half-life (greater than 1 month) and low activity in nuclear physics, nuclear astrophysics, and nuclear technological measurements.
This work demonstrates that JURLab is well suited for measuring low activities, whether arising from the long lifetime of the decaying nucleus or from the small cross section of the nuclear reaction. Building on the first activation cross section measurement performed at the facility~\cite{Szegedi_2021idx}, we plan to carry out similar studies relevant to nuclear astrophysics or nuclear technology relevance.

\section{Characterization of the laboratory}

In August 2021, a high-precision experimental program began at the HUN-REN Wigner Research Centre for Physics' Jánossy Underground Research Laboratory to study possible nuclear decay-rate anomalies~\cite{PoS_Fenyvesi}. This laboratory is a shallow underground, low-radiation-background site at Csillebérc, Budapest, Hungary.

Figure~\ref{timeline} shows the measurement timeline over different data-taking periods.
The data acquisition was divided into continuous data-taking periods (Runs) listed in Table~\ref{Tbl_runs}. The data were collected over multiple runs rather than in a single uninterrupted acquisition period. Each run corresponds to a fixed detector geometry; any modification to the detector configuration marked the start of a new run. In a few instances, the radioactive source was temporarily removed for administrative purposes, such as inspections. Technical interruptions were rare, brief, and did not alter the detector geometry.

Shortly after an activation thick-target-yield measurement~\cite{Szegedi_2021idx}, we started the Run-0 test period from July 7, 2022 to October 7, 2022 involving a long background measurement in JURLab. The test period was followed by Runs~1-4 during which a $^{137}$Cs source was placed in front of the detector. The aim of these runs was to check detector stability, test the electronics and optimize the settings.
Run-1 lasted from October 18, 2022 to March 3, 2023 with a very short break in data acquisition on March 12, 2023. For Run-2 the \textsuperscript{137}Cs source was moved closer to the detector resulting in an increase in total counts detected visible on Figure~\ref{timeline}; the run operated without any interruptions from May 1, 2023 to August 9 2023. Run-3 ran from September 1, 2023 to January 25, 2024 with a break from December 10, 2023 to 24 December 2023 when an impulse generator was connected and the \textsuperscript{137}Cs isotope was removed for mandatory inspection resulting in a drop in total counts shown on Figure~\ref{timeline}; since the source was placed back to its original position after inspection, data acquisition of Run-3 resumed smoothly. Run-4 operated from January 25, 2024 to April 10, 2024 without any interruptions.

\begin{table*}[htbp]
\centering
\caption{Run periods of the measurement at JURLab between 2022 and 2024.}
\begin{tabular*}{\tblwidth}{@{}LRRL@{}}
\toprule
Run-\# & Start date & End date & Remark \\
\midrule
Run-0 & 2022-07-07 & 2022-10-07 & Test run: radioisotope positioning \\
Run-1 & 2022-10-18 & 2023-03-24 & Short DAQ system interruption \\
Run-2 & 2023-05-01 & 2023-08-09 & Source moved closer to detector \\
Run-3 & 2023-09-01 & 2024-01-25 & Pause: signal generator connected \\
& & & and isotope removed for inspection \\
Run-4 & 2024-01-25 & 2024-04-10 & Experiment stopped; isotope removed \\
\bottomrule
\end{tabular*}
\label{Tbl_runs}
\end{table*}

\subsection{Jánossy Underground Research Laboratory}

\begin{figure}[h!]
  \centering
  \includegraphics[width=0.4\textwidth]{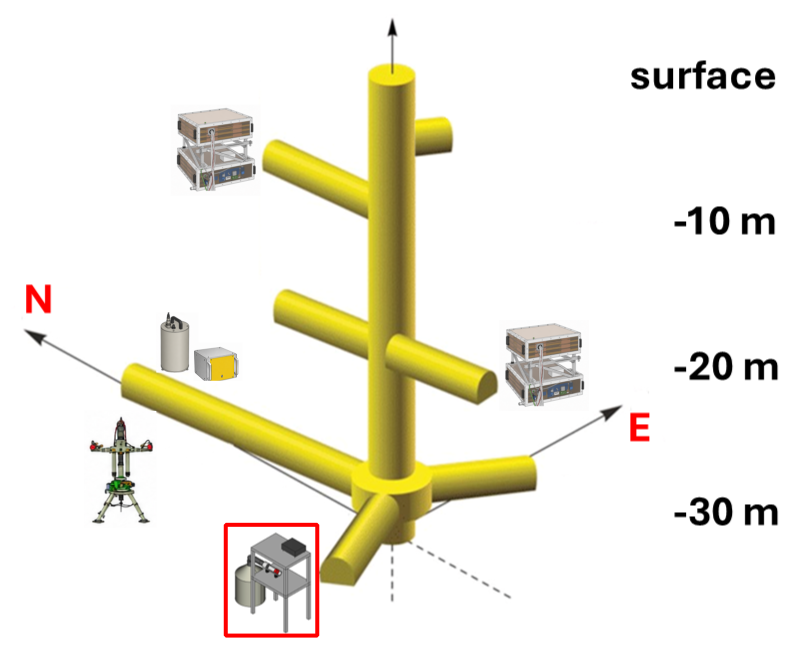}
  \includegraphics[width=0.4\textwidth]{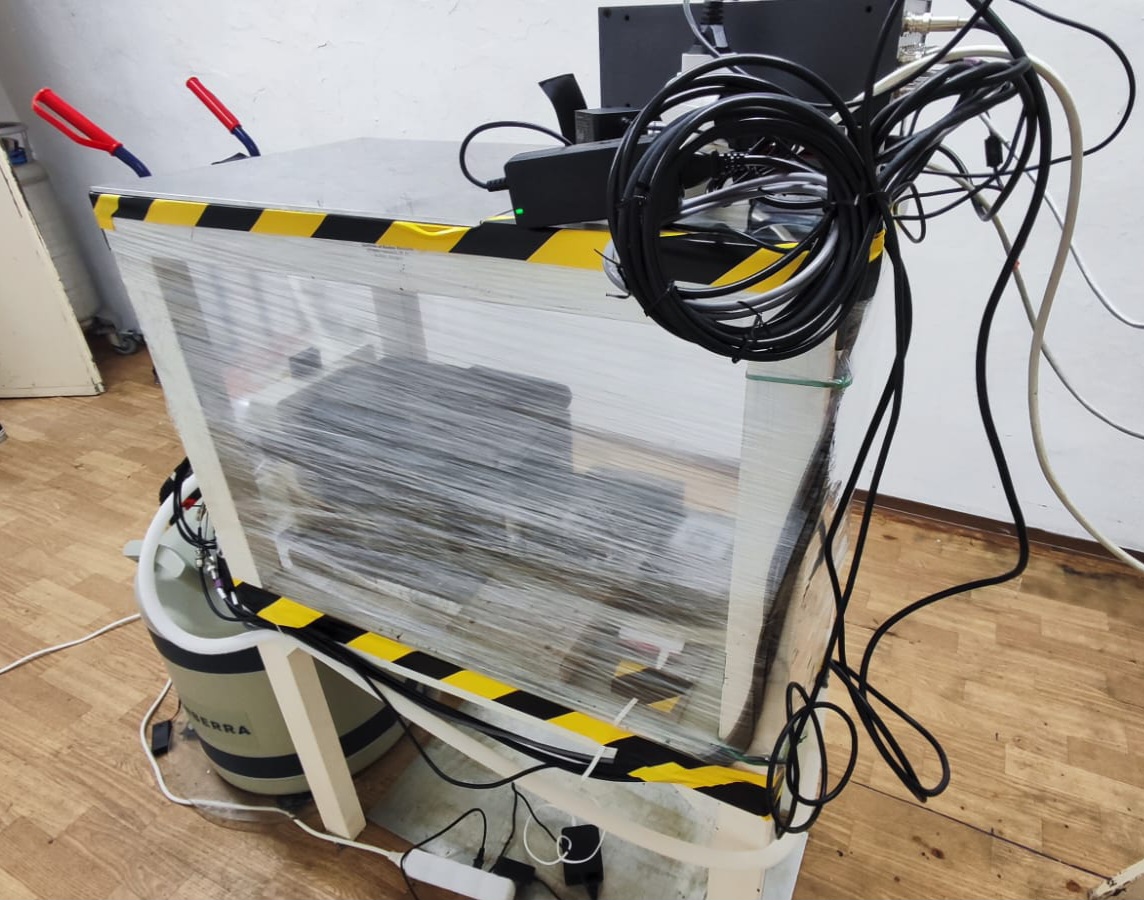}
  \caption{Setup of measurements conducted at HUN-REN Wigner RCP's Jánossy Underground Research Laboratory (JURLab). The low-background $\gamma$-counting facility is located at $-30$~m (top panel framed in red). The other experiments running parallel are: muon tomographs are located at Level~$-1$, $-2$, and at Level $-3$ in an another hall an Eötvös balance and a Güralp seismometer are mounted. HPGe detector with its shielding is shown (bottom panel).}
  \label{fig:JURLAB_lab}
\end{figure}

The Jánossy Underground Research Laboratory is a special shallow underground laboratory surrounded by Dach\-stein-type limestone. It has 40~cm-thick outer walls made of reactor-grade concrete, making the site suitable for measurements requiring low cosmic-ray background. It has three levels below ground:
a room of 20~m$^2$ at a depth of 10~m (Level~$-1$),
two halls (2~$\times$~20~m$^2$) at a depth of 20~m (Level~$-2$),
and three halls (2~$\times$~20~m$^2$ and 50~m$^2$) at a depth of 30~m (Level~$-3$) shown in Figure~\ref{fig:JURLAB_lab}. The low-background $\gamma$-counting site is located at the lowest level of JURLab, corresponding to an overburden of approximately 75~m.w.e.~\cite{OLAH2014_clean}. This situates the site's depth between the CAVE laboratory of the International Atomic Energy Agency IAEA in Monaco (54~m.w.e.)~\cite{Laubenstein2004Underground} 
and bunker~110 of the Felsenkeller laboratory in Dresden, Germany (140~m.w.e.)~\cite{TURKAT2023102816}. At Level~$-3$ in JURLab, the cosmic-ray muon flux is reduced by more than one order of magnitude compared to surface-level measurements.

At JURLab an uninterrupted power supply (UPS) unit provides electricity for the instruments of the experiments operating in the laboratory. A Heating, Ventilation, and Air Conditioning (HVAC) system ventilates the air at the facility. Under normal operation, it runs for a few consecutive hours each week on Monday to reduce humidity and radon levels in the laboratory. The temperature can be kept constant to within $\pm 0.2^{\circ}$C. If the HVAC system is operated continuously, the relative humidity remains constant to within $\pm 5\%$. This allows the radon concentration to be reduced and kept almost constant through air circulation.

The following measurements are currently being carried out in the laboratory: low-radiation background measurements with a Canberra High Purity Germanium (HPGe) detector~\cite{PoS_Fenyvesi}, muon tomography measurements with muon detectors~\cite{VARGA2020162236}, a remeasurement of the Eötvös experiment with a modernized Eötvös balance~\cite{Ebalance,Völgyesi:20206K}, and seismic monitoring with a Güralp seismometer and a Raspberry Shake 4D. A tiltmeter and an infrasound monitoring system are also in operation at the site. Since multiple experiments are running in parallel at JURLab, remote control is important in order to minimize human presence and environmental disturbance.

\section{Environmental monitoring and $\gamma$-background at JURLab}

The centerpiece of the low-background counting facility is a Canberra GC4018 type HPGe detector~\cite{Canberra2024}. The used device is a $p$-type coaxial detector with excellent energy resolution in the photon-energy range of interest, $ 40~\mathrm{keV} \lesssim E_{\gamma} \lesssim 10$~MeV, and good relative efficiency ($\gtrsim 40\%$). Its full width at half maximum is $\lesssim 0.87$~keV and $\lesssim 1.80$~keV at 122~keV and 1.33~MeV, respectively, and its peak-to-Compton ratio is $\gtrsim 62$. The detector's Dewar stores enough liquid nitrogen for about a week of operation. Additionally, lead shielding of a few centimeters of thickness is built around the detector to further suppress the $\gamma$-background.
The data-acquisition system (DAQ) is a Canberra Lynx digital signal analyzer~\cite{Lynx2024}. It has up to 32k channels, accurate time stamps, excellent count-rate and temperature stability, and Ethernet connectivity. Data acquisition display and control, system energy and shape calibration, and operation of the digital oscilloscope are performed via the HTTP interface of the Lynx multichannel analyzer. Spectral display, acquisition control, energy calibration, and instrument setup are also possible through the built-in web server. Thus, apart from liquid-nitrogen refills, the experiment can be run remotely with essentially no human presence.
The primary data collected by the Lynx analyzer are transferred to the HUN-REN Wigner Datacenter and the HUN-REN Wigner Scientific Computing Laboratory (WSCLAB), allowing continuous monitoring and analysis while the experiment is running.

\begin{figure}[h!]
  \centering
  \includegraphics[width=\columnwidth]{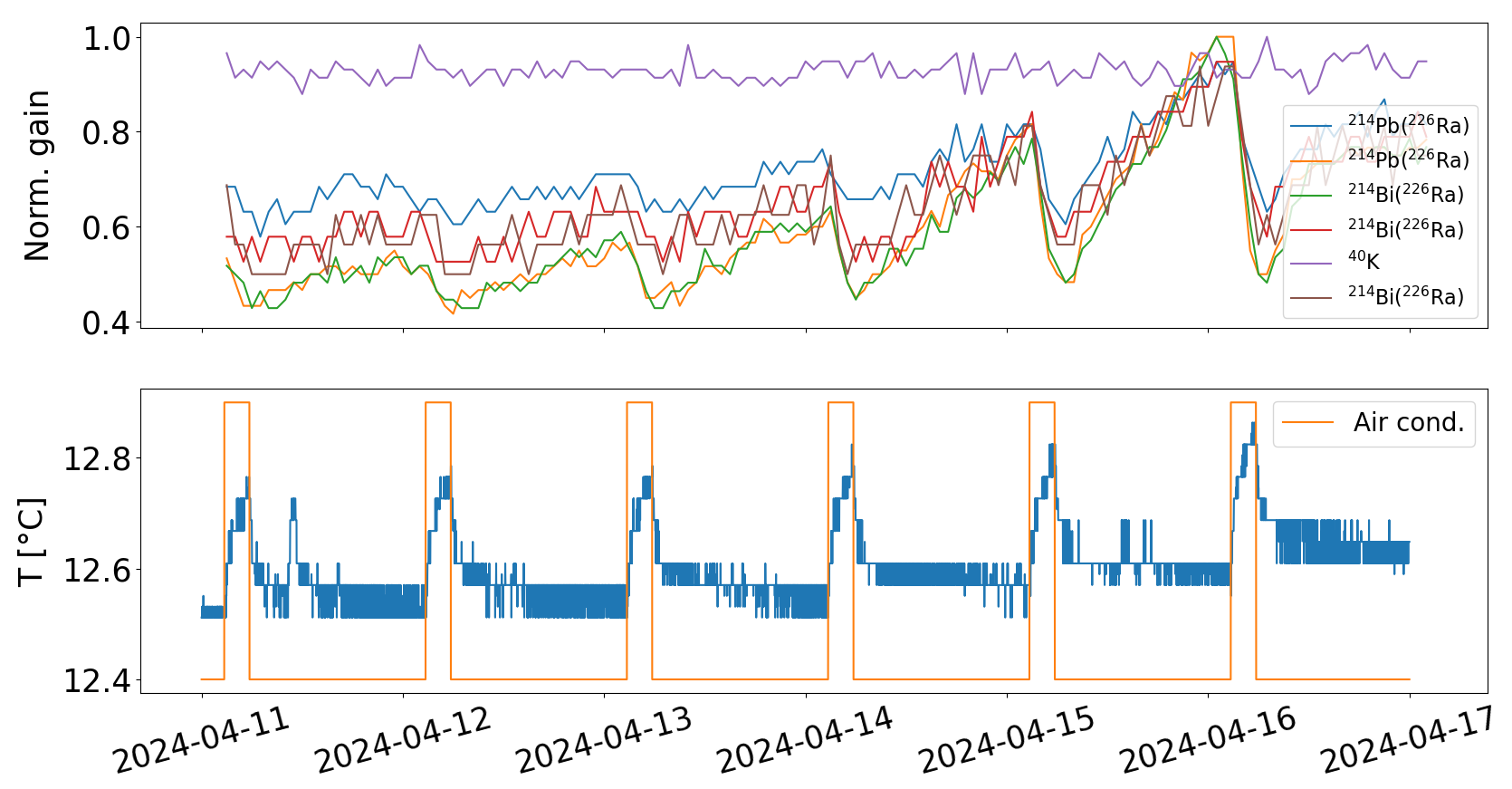}
  \caption{Relative activities of the daughters of \textsuperscript{222}Rn and the laboratory temperature $T$ with the HVAC system at Level~$-3$ of the JURLab shown on the top and bottom panels, respectively.}
  \label{outside_environment}
\end{figure}

Temperature and relative humidity near the HPGe detector are monitored by an AM2315 sensor connected to a Raspberry Pi computer used for periodic sensor readout; the results are then transferred via Ethernet for storage and analysis. The humidity and temperature sensors of the HVAC system are also monitored remotely via Ethernet. The temperature at the lowest level of JURLab remained between 12.2$^{\circ}$C and 12.7$^{\circ}$C for the duration of all runs.
Figure~\ref{outside_environment} shows the relationship between the activities of the daughters of \textsuperscript{222}Rn and the variation of environmental parameters, in particular the laboratory temperature $T$ over a period of one week during Run-4 with the air-conditioning system operating for one hour each day for these runs specifically. The activities of the isotopes shown in the top panel were normalized to unity. It is visible that the daughters of \textsuperscript{222}Rn were effectively reduced by the air-conditioning system, while the activity of \textsuperscript{40}K (top panel, purple line) remained within a narrow band. This behavior coincides with the on/off state of the HVAC system (bottom panel, yellow line). A similar correlation is visible in the laboratory temperature $T$ (bottom panel, blue line), which increases slightly while the HVAC system is on and relaxes back to the ambient underground temperature after the system is turned off.

 \begin{figure}[h!]
  \centering
  \includegraphics[width=\columnwidth]{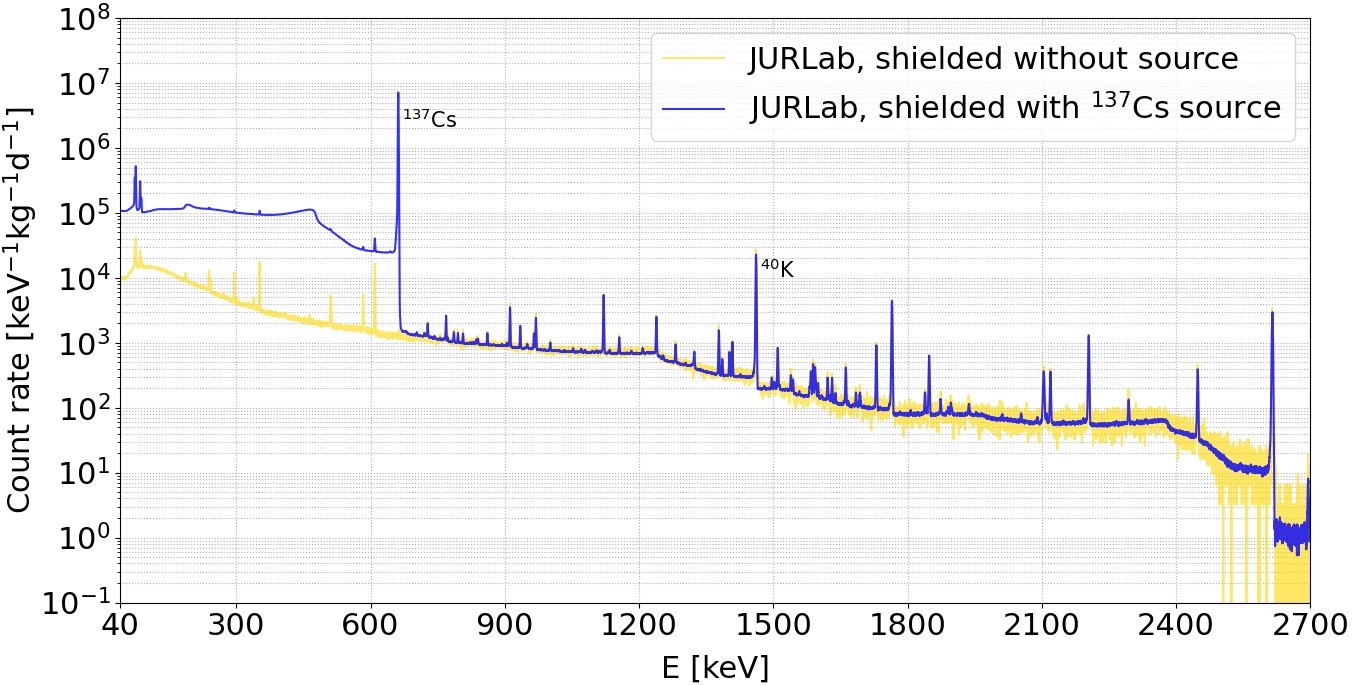}
  \caption{The dark blue line represents the measured \textsuperscript{137}Cs gamma-ray spectra for Run-1 with 2.5~cm lead shielding and the yellow line shows the background without the source.} 
  \label{Cs_spectrum}
\end{figure}
\begin{figure*}[h!]
  \centering
  \includegraphics[width=1.5\columnwidth]{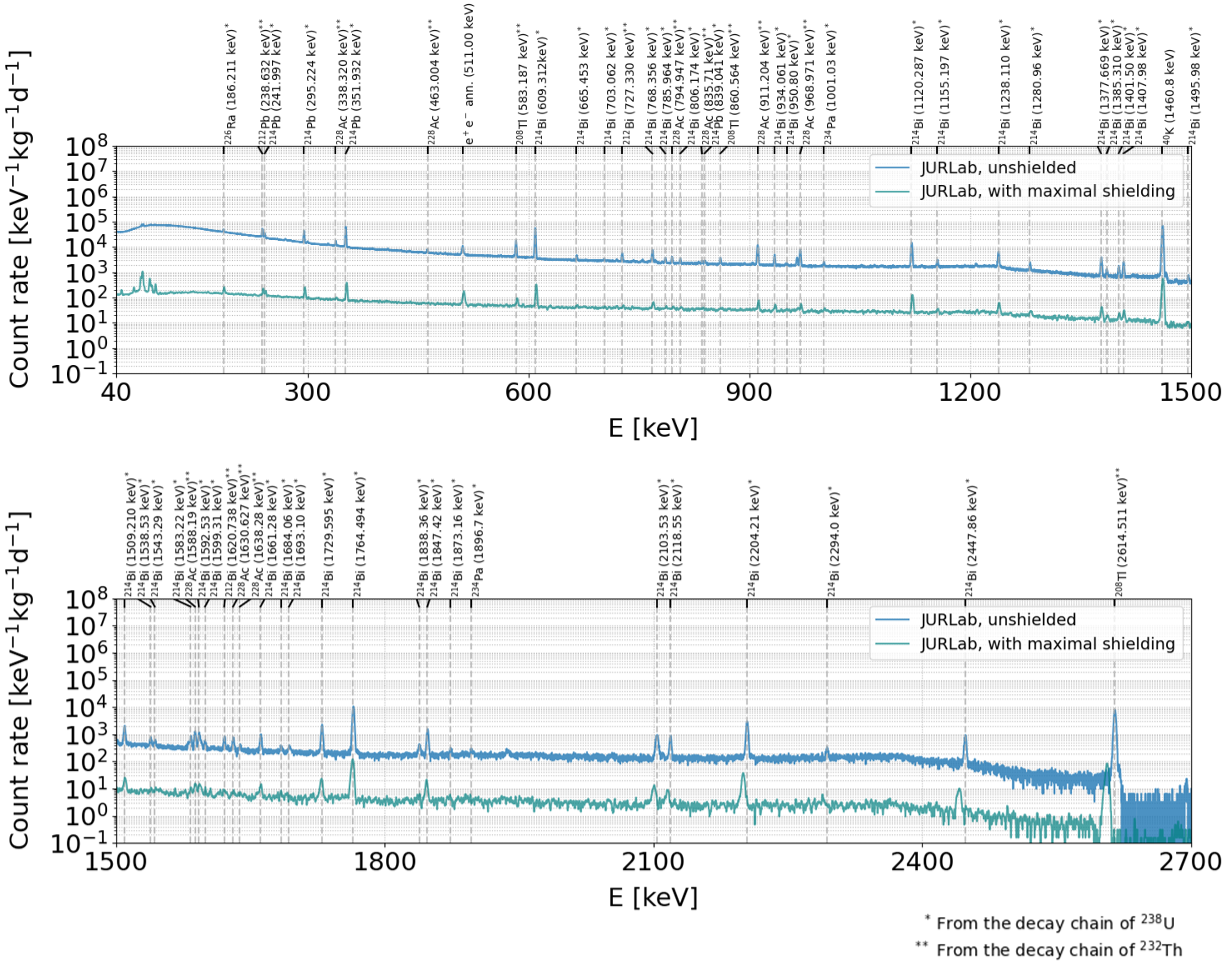}
  \caption{Gamma-ray spectra measured at Level~$-3$ of JURLab. The blue line shows the background without shielding, while the dark green line represents the background with 5-10~cm lead and 2~mm copper shielding.}
  \label{background_spectrum}
\end{figure*}

As an example Figure~\ref{Cs_spectrum} shows the Run-1 measured spectra with the underlying 2.5~cm lead shielded background with and without the \textsuperscript{137}Cs source present. The \textsuperscript{137}Cs and \textsuperscript{40}K peaks at 661.657~keV and 1460.820~keV, respectively, stand out significantly above the background.
While Figure~\ref{background_spectrum} presents background spectra measured with and without the 5-10~cm lead and 2~mm copper shielding normalized to the counting length and bin size.

\begin{figure}[h!]
    \centering
    \includegraphics[width=\columnwidth]{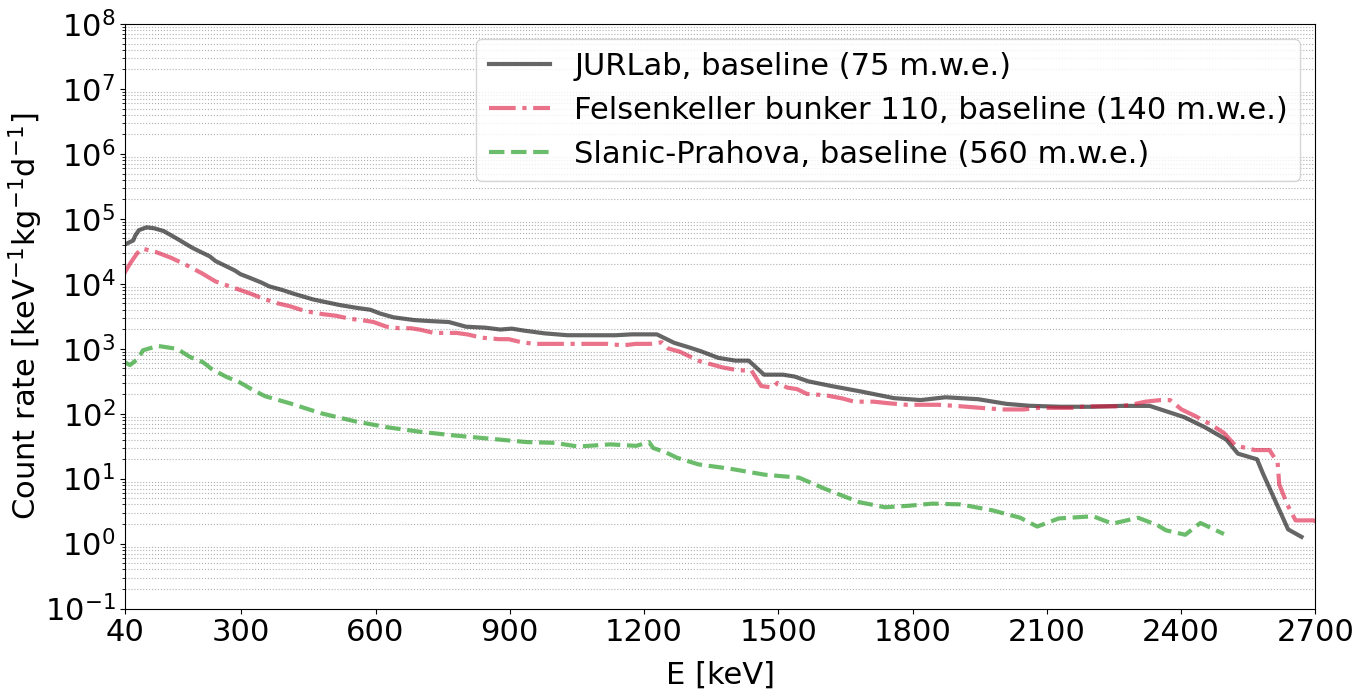}
    \caption{Unshielded background baselines at JURLab (75 m.w.e.), Felsenkeller bunker 110 (140 m.w.e.)~\cite{TURKAT2023102816} and Sl\u{a}nic Prahova (560 m.w.e.)~\cite{MARGINEANU20081501} shown for approximate comparison in black solid, red dashed-dotted and green dashed lines, respectively.}
    \label{fig:compare_baseline}
\end{figure}

Figure~\ref{fig:compare_baseline} places the present measurement of the unshielded JURLab background in the context of published indicative baseline spectra of other underground laboratories. The comparison includes the JURLab (75 m.w.e.) baseline extracted from the measured unshielded spectrum with the baselines of the Felsenkeller bunker~110 (140 m.w.e.)~\cite{TURKAT2023102816} and the Sl\u{a}nic Prahova site (560 m.w.e.)~\cite{MARGINEANU20081501}. These curves were approximated by smooth fits to the spectra reported. Detector type, geometry, shielding, and normalization all differ for each laboratory therefore the comparisons are intended only as approximate benchmarks rather than re-analyzes. The comparison illustrates the decrease in background with increasing underground overburden. 

\begin{figure}[h!]
    \centering
    \includegraphics[width=\columnwidth]{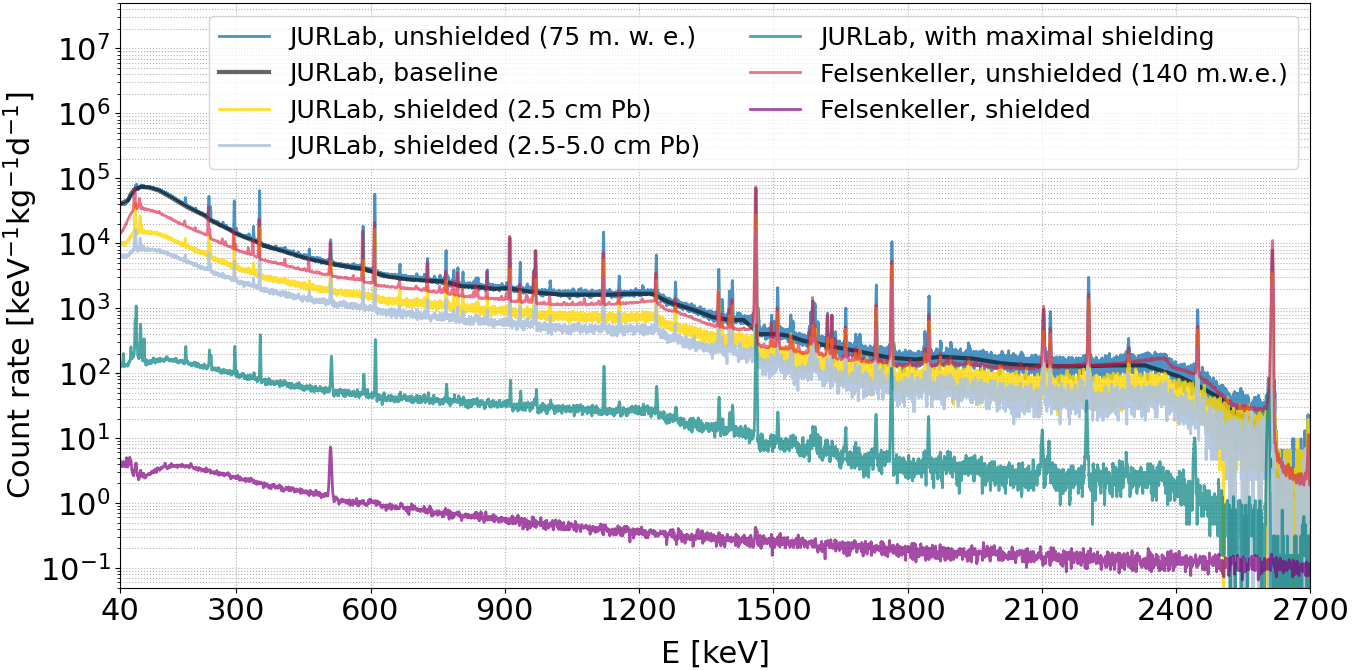}
    \caption{Background spectra measured at JURLab with corresponding data reported for Felsenkeller bunker 110~\cite{TURKAT2023102816}. (See text for details.)}
    \label{fig:compare_spectra}
\end{figure}

Figure~\ref{fig:compare_spectra} shows the reduction in background levels at JURLab achieved through improvements to the detector shielding and compares these measurements with corresponding spectra from Felsenkeller bunker 110. In addition to the unshielded spectrum shown by the blue curve, the JURLab spectra also include the initial 2.5~cm lead shielding spectrum measured during Runs 1–4 shown by the yellow curve. The silver curve shows the background after upgrading the shielding to 2.5–5~cm lead together with active radon suppression by continuous flushing. Finally, the dark green curve corresponds to the present best-achieved configuration, employing 5-10~cm full $4\pi$ lead shielding together with an additional 2~mm copper inner liner. Although these optimized configurations were implemented after the data analyzed in this work were collected, they are included to demonstrate the expected background level available for future decay-rate measurements. 
At Felsenkeller, the application of 15~cm lead and 10~cm copper shielding shown by the red curve reduces the background by approximately four orders of magnitude relative to the bunker 110's unshielded spectrum indicated by the purple curve. While at JURLab, the 5-10~cm lead and 2~mm copper shielding reduces the background by approximately two orders of magnitude relative to the unshielded measurement. Although shielding configurations and detector systems differ and therefore do not permit a direct performance comparison, these results demonstrate the substantial background reduction obtainable through local shielding. Further improvements, including additional shielding and a cosmic-ray veto, could provide further suppression of the residual background.

\section{Investigating the detector stability over a long counting period}

When measuring low activities, it is critical not only to keep the laboratory background low but also to ensure the stability of the detection setup over time. This applies both to the detection efficiency and to the stability of all elements of the electronic chain. In order to examine the stability of the system (i.e. efficiency, energy calibration, etc.), a $^{137}$Cs source was placed at a distance of 14.5 cm from the detector's surface and the detector stability was evaluated using the ratio of the \textsuperscript{137}Cs and \textsuperscript{40}K photopeak areas. Since both peaks were recorded simultaneously, they were subject to the same environmental and instrumental conditions. The photopeak of \textsuperscript{40}K serves as a stable reference because of its long half-life ($T_{1/2} = 1.2522(27)\times10^{9}$ years~\cite{Mougeot_2026}) and as a constant content of the rock surrounding the laboratory.

\begin{figure}[h!]
    \centering
    \includegraphics[width=0.48\textwidth]{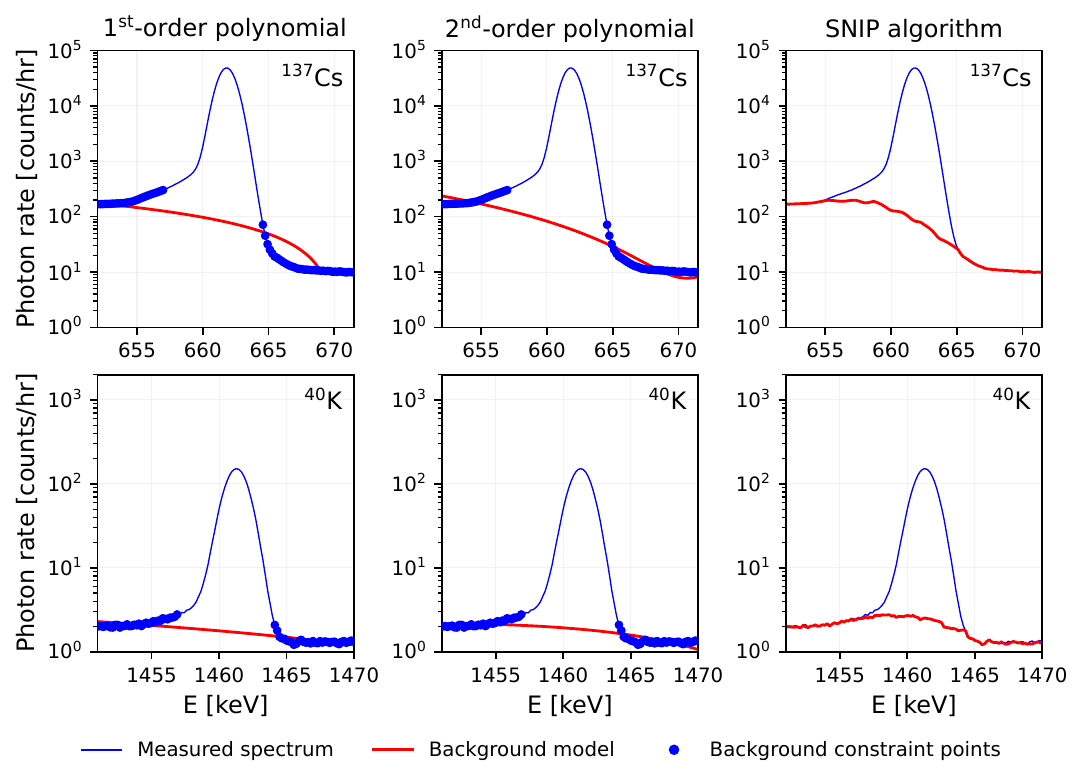}
    \caption{Comparison of three background-subtraction methods applied to the \textsuperscript{137}Cs (top row) and \textsuperscript{40}K (bottom row) photopeaks. From left to right, the background is modeled using a 1\textsuperscript{st}-order polynomial, a 2\textsuperscript{nd}-order polynomial, and the SNIP algorithm~\cite{He2015_08, SNIP1988}. Blue markers indicate the background constraint points (sidebands) used in the polynomial fits; the SNIP method is independent of explicit sideband regions.}
    \label{background_subtraction}
\end{figure}

The background under the \textsuperscript{137}Cs and \textsuperscript{40}K photopeaks in the measured raw spectrum was estimated using the defined background regions [652.0 - 657.0]~keV, [664.5 - 671.5]~keV and [1451.0 - 1457.0]~keV, [1464.0 - 1470.0]~keV, respectively. After peak extraction, an analogous background-subtraction procedure was applied, and finally the total peak area was determined by direct summation of background-subtracted counts. Figure~\ref{background_subtraction} shows the three different back\-ground-modeling procedures implemented: 1\textsuperscript{st}-order polynomial fit, 2\textsuperscript{nd}-order polynomial fit to the low- and high energy background regions, and use of the Statistics-sensitive Non-linear Iterative Peak-clipping (SNIP) algorithm~\cite{He2015_08, SNIP1988} which estimates the slowly varying background directly from the spectrum without requiring predefined background regions. To test the robustness of the background subtraction, we studied the variation in the extracted peak area when the integration range was progressively narrowed by up to 30\%, i.e. to [653.5 - 670.5]~keV, and found less than a 0.1\% relative change in the peak area in all cases. The extracted peak areas obtained using the three background-subtraction methods show good overall consistency. 
For the \textsuperscript{137}Cs peak, the background-corrected peak areas obtained with 1\textsuperscript{st}-order polynomial, 2\textsuperscript{nd}-order polynomial, and SNIP background modeling agree within approximately 0.6\%. While, for the \textsuperscript{40}K peak, the corresponding relative spread is slightly larger, ranging from 1.6\% to 2\%.

\begin{figure}[h!]
    \centering
    \includegraphics[width=0.45\textwidth]{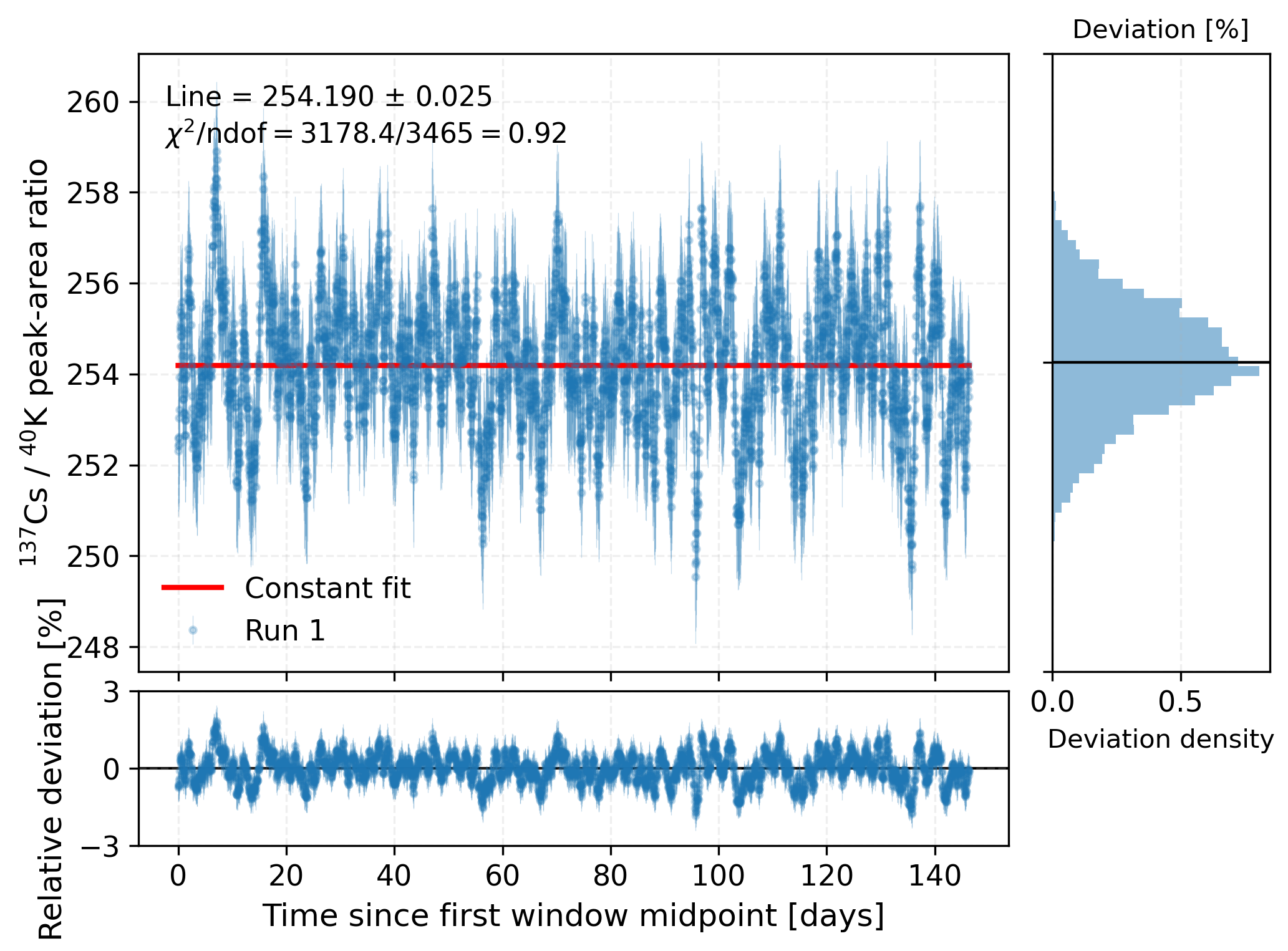}
    \caption{Corrected $^{137}$Cs/$^{40}$K photopeak-area ratio as a function of time for Run-1. The ratio was calculated using fixed integration windows and 24-hour sliding-window averaging. The red line shows a linear fit to the data, used to characterize residual long-term trends in the ratio. The bottom panel shows the relative deviation, while the right panel shows the deviation density of the data.} 
    \label{fig:cs_k_ratio}
\end{figure}

For the analysis, the dead-time values provided and stored automatically in the recorded CNF data file by the Lynx analyzer were used and live-time corrected. The background-subtracted peak areas were corrected for dead-time by scaling them to the nominal one-hour measurement duration by a factor $\Delta t_{\rm nominal}/\Delta t_{\rm live}$.
The spectra were grouped into sliding windows of 24 consecutive one-hour acquisitions. Net peak areas were extracted using the previously defined background regions and a linear background subtraction. The \textsuperscript{137}Cs peak areas were corrected for radioactive decay, while no correction was applied to \textsuperscript{40}K. The resulting ratio,
\begin{equation}
R(t) = \frac{A_{^{137}\mathrm{Cs}}(t)}{A_{^{40}\mathrm{K}}(t)},
\end{equation}
was evaluated as a function of time for each run. Figure \ref{fig:cs_k_ratio} shows the decay-corrected \textsuperscript{137}Cs/\textsuperscript{40}K photopeak-area ratio as a function of time for Run-1. The measured data was fitted with a linear function, its slope is found to be always consistent with zero within uncertainties which demonstrates the long-term stability of the system. We also examined the stability of the electronic settings. The centroid of the curves fitted to the transitions corresponding to the $^{40}$K and $^{137}$Cs isotopes changed almost negligibly (approximately 10$^{-4}$ keV/day) during the counting. 

As part of the laboratory characterization, we provide an estimated detection limit (minimum detectable activity)~\footnote{The detection limit is defined as $L_D=k^2+2k(2B)^{1/2}$, where $B$ is the estimated background below the peak and k corresponds to the confidence level ($k=1.645$ is equivalent to 95\% confidence level)~\cite{Leo1987}.}. The Budapest Neutron Centre (BNC), a research reactor of the HUN-REN Energy Research Institute located on the KFKI campus near JURLab, performs (n, $\gamma$) activation measurements in combination with $\gamma$ spectroscopy. Furthermore, the HUN-REN Institute for Nuclear Research (ATOMKI, located in Debrecen about 200~km from Budapest) activation analysis is also performed using the 2~MV Tandetron and $K=20$ cyclotron accelerators. Accordingly, such samples can also be studied at JURLab, where isotopes with half-lives of a few hours or days can be measured following production at the nearby reactor or at the accelerator facility, respectively. 

\begin{table*}[h!]
\centering
\caption{Preliminary minimum detectable activities ($A_{\mathrm{min}}$) for
selected radionuclides calculated using the shielded background spectrum.}
\label{tab:candidates}
\begin{tabular*}{\tblwidth}{@{}LLLLLLL@{}}
\toprule
Nuclide & Half-life & $E_\gamma$ & $I_\gamma$ & ROI &
$L_D$  & $A_{\mathrm{min}}$  \\
 & &  (keV) & & (keV) & (day$^{-1}$)& (Bq)\\
\midrule
$^{22}$Na  & 2.603 years & 1274.54 & 0.9994 & 1269--1280 & 304 & 4.99  \\
$^{103}$Ru & 39.25 days  & 497.08  & 0.9080 & 493--501   & 492 & 4.42  \\
$^{145}$Sm & 340 days    & 61.23   & 0.1215 & 59--63     & 821 & 33.24 \\
$^{65}$Ni  & 2.52 hours  & 1481.84 & 0.236  & 1476--1488 & 205 & 16.00 \\
$^{97}$Zr  & 16.75 hours & 743.36  & 0.931  & 739--748   & 405 & 4.78  \\
$^{7}$Be   & 53.22 days  & 477.60  & 0.1044 & 474--481   & 480 & 36.41 \\
$^{95}$Zr  & 64.0 days   & 756.73  & 0.544  & 752--761   & 397 & 8.12  \\
\bottomrule
\end{tabular*}
\end{table*}

Table~\ref{tab:candidates} summarizes candidate radionuclides with half-lives ranging from a few hours to a few weeks, selected according to three criteria: a well-isolated, intense $\gamma$ line, clear motivation in nuclear astrophysics or nuclear technology, and low induced activity. The radionuclides' energy and intensity of the emitted radiation, as well as the minimum activity detectable at JURLab are derived. One of the isotopes mentioned as an example is $^{22}$Na. The measurement of the $^{19}$F($\alpha$,n)$^{22}$Na reaction, introduced earlier, is currently in progress at HUN-REN ATOMKI. The second example is the $^{103}$Ru nucleus. The minimum detectable activity was determined to design the irradiations required to measure the $^{100}$Mo($\alpha$,n)$^{103}$Ru reaction in the astrophysically relevant energy range~\cite{Szegedi_2021idx}. The third example is related to the study of the $^{142}$Nd($\alpha$,n)$^{145}$Sm reaction, where the goal is to investigate the optical potentials used in the modeling of the astrophysical $\gamma$-process to measure the reaction cross section over a wide energy range.
Further candidates to probe the main and weak $s$-process branchings and neutron-density monitors, produced via (n,$\gamma$) reactions, are e.g. $^{95,97}$Zr $^{65}$Ni or $^{153}$Sm~\cite{Kappeler2011,Bisterzo2015,Mosconi2010,Abbondanno2004,Reifarth2003,Pignatari2010}. Astrophysically relevant reactions resulting $^{7}$Be are also candidates for further experimental investigations \cite{Tot23, Tot24}.

\section{Discussion and conclusion}

We presented the experimental setup and initial characterization of a long-term \textsuperscript{137}Cs decay-rate measurement performed at the Jánossy Underground Research Laboratory of the HUN-REN Wigner Research Centre for Physics. The underground location provides a low-back\-ground environment and stable operating conditions that are advantageous for precision studies of radioactive decay.

Using data collected between the period of October~2022 and April~2024, corresponding to approximately 13 months of measurement time, we assessed the stability of the detector response through the \textsuperscript{137}Cs and \textsuperscript{40}K photopeak areas and, in particular, their ratio as well as the suitability of the setup for long-duration measurements. As a complementary diagnostic, we also examined the centroid evolution of both photopeaks within each run. This shows that the calibration drift over time is almost negligible. The JURLab background measurements were compared with published baseline spectra from underground laboratories: Felsenkeller bunker 110 and Sl\u{a}nic Prahova. Comparison with published underground background spectra shows that the present JURLab performance is comparable to that of other shallow underground laboratories, while further improvements in passive shielding and the planned implementation of an active muon-veto system are expected to reduce the residual background further.

In summary, this work establishes the experimental infrastructure and identifies the main systematic limitations of the current implementation. As such, it provides the methodological foundation for future analyses aimed at substantially improving the stability and sensitivity of long-term decay-rate and low-activity measurements conducted at JURLab.

\section*{Acknowledgements}

This work was supported by the FuSe COST Action CA24101, the HUN-REN Office of Supported Research Groups (TKI) under award No. TKCS-2024/56 (GGK), and by the Hungarian National Research, Development and Innovation Office (NKFIH) under the Contracts NKFIH K147010 (GGK), NKKP ADVANCED\_24 150038 (EF) and ADVANCED\_25 K153456 (GGB), 2024-1.2.5-TÉT-2024-00022 (GGB), 2025-1.1.5-NEMZ\_KI-2025-00005 (GGB), and 2025-1.1.5-NEMZ\_KI-2025-00011 (EF).

Instrument R\&D and operation were carried out at the Jánossy Underground Research Laboratory (JURLab) of the Vesztergombi Laboratory for High Energy Physics (VLAB), the Wigner Datacenter, and the Wigner Scientific Computing Laboratory (WSCLAB) at the HUN-REN Wigner Research Centre for Physics.

\FloatBarrier

\bibliographystyle{cas-model2-names}
\bibliography{cas-refs}

@article{ALBURGER1986168,
title = {Half-life of \textsuperscript{32}{Si}},
journal = {Earth and Planetary Science Letters},
volume = {78},
number = {2},
pages = {168-176},
year = {1986},
issn = {0012-821X},
doi = {https://doi.org/10.1016/0012-821X(86)90058-0},
url = {https://www.sciencedirect.com/science/article/pii/0012821X86900580},
author = {D.E. Alburger and G. Harbottle and E.F. Norton}
}

@incollection{OLAH2014_clean,
  author    = {L. Oláh and G. G. Barnaföldi and G. Hamar and H. G. Melegh and G. Surányi and D. Varga},
  title     = {Applications of Cosmic Muon Tracking at Shallow Depth Underground},
  booktitle = {Astroparticle, Particle, Space Physics and Detectors for Physics Applications},
  pages     = {280--284},
  year      = {2014},
  publisher = {World Scientific},
  doi       = {10.1142/9789814603164\_0043}
}

@article{TURKAT2023102816,
title = {A new ultra low-level HPGe activity counting setup in the Felsenkeller shallow-underground laboratory},
journal = {Astroparticle Physics},
volume = {148},
pages = {102816},
year = {2023},
issn = {0927-6505},
doi = {https://doi.org/10.1016/j.astropartphys.2023.102816},
url = {https://www.sciencedirect.com/science/article/pii/S0927650523000026},
author = {S. Turkat and D. Bemmerer and A. Boeltzig and A.R. Domula and J. Koch and T. Lossin and M. Osswald and K. Schmidt and K. Zuber}
}

@article{Laubenstein2004Underground,
  title = {Underground measurements of radioactivity},
journal = {Applied Radiation and Isotopes},
volume = {61},
number = {2},
pages = {167-172},
year = {2004},
note = {Low Level Radionuclide Measurement Techniques - ICRM},
issn = {0969-8043},
doi = {https://doi.org/10.1016/j.apradiso.2004.03.039},
author = {M Laubenstein and M Hult and J Gasparro and D Arnold and S Neumaier and G Heusser and M Köhler and P Povinec and J.-L Reyss and M Schwaiger and P Theodórsson}
}

@article{VARGA2020162236,
title = {Detector developments for high performance Muography applications},
journal = {Nuclear Instruments and Methods in Physics Research Section A: Accelerators, Spectrometers, Detectors and Associated Equipment},
volume = {958},
pages = {162236},
year = {2020},
note = {Proceedings of the Vienna Conference on Instrumentation 2019},
issn = {0168-9002},
doi = {https://doi.org/10.1016/j.nima.2019.05.077},
url = {https://www.sciencedirect.com/science/article/pii/S0168900219307417},
author = {D. Varga and G. Nyitrai and G. Hamar and G. Galgóczi and L. Oláh and H.K.M. Tanaka and T. Ohminato}
}

@misc{Ebalance,
  author = {L. Völgyesi and G.G. Barnaföldi and Cs. Égető and E. Fenyvesi and B. Kiss and P. Lévai and Gy. Szondy and Gy. Tóth and P. Ván},
  title = {Current use of {E}\"otv\"os torsion balance, the tidal effect},
  year = 2024,
  url = {https://www.issmge.org/uploads/publications/25/111/ISC2020-519.pdf},
  howpublished = {\url{https://www.issmge.org/uploads/publications/25/111/ISC2020-519.pdf}},
  urldate = {2024-06-12}
}

@inproceedings{Völgyesi:20206K,
  author = "Völgyesi, Lajos  and  Szondy, György  and  Tóth, Gyula  and  Péter, Gábor  and  Kiss, Bálint  and  Barnaföldi, Gergely Gábor  and  Deák, László  and  Égető, Csaba  and  Fenyvesi, Edit  and  Gróf, Gyula  and  Somlai, László  and  Harangozó, Péter  and  Levai, Peter  and  Ván, Péter",
  title = "{Preparations for the remeasurement of the Eötvös Experiment}",
  doi = "10.22323/1.353.0041",
  booktitle = "Proceedings of International Conference on Precision Physics and Fundamental Physical Constants {\textemdash} PoS(FFK2019)",
  year = 2020,
  volume = "353",
  pages = "041"
}

@article{Ellis_1990,
doi = {10.1088/0031-9155/35/8/004},
url = {https://dx.doi.org/10.1088/0031-9155/35/8/004},
year = {1990},
month = {aug},
publisher = {},
volume = {35},
number = {8},
pages = {1079},
author = {K J Ellis},
title = {The effective half-life of a broad beam \textsuperscript{238}{Pu}/{Be} total body neutron irradiator},
journal = {Physics in Medicine \& Biology}
}

@article{Falkenberg2001,
	author = {Eckhard Dieter Falkenberg},
	journal = {Apeiron},
	number = {2},
	pages = {32--45},
	title = {Radioactive Decay Caused by Neutrinos},
	volume = {8},
	year = {2001}
}

@article{Parkhomov,
doi = {10.4236/jmp.2011.211162},
url = {},
issn = {},
year = {2011},
month = {},
day={},
publisher = {},
volume = {2},
number = {11},
pages = {1310-1317},
author = {A. Parkhomov},
title = {Deviations from Beta Radioactivity Exponential Drop},
journal = {Journal of Modern Physics}
}

@article{Sturrock2011,
doi = {10.1007/s11207-011-9807-5},
url = {https://doi.org/10.1007/s11207-011-9807-5},
issn = {1573-093X},
year = {2011},
month = {07},
day={06},
publisher = {},
volume = {272},
number = {1},
pages = {1},
author = {P. A. Sturrock and E. Fischbach and J. H. Jenkins},
title = {Further Evidence Suggestive of a Solar Influence on Nuclear Decay Rates},
journal = {Solar Physics}
}

@article{JENKINS2009407,
title = {Perturbation of nuclear decay rates during the solar flare of 2006 December 13},
journal = {Astroparticle Physics},
volume = {31},
number = {6},
pages = {407-411},
year = {2009},
issn = {0927-6505},
doi = {https://doi.org/10.1016/j.astropartphys.2009.04.005},
url = {https://www.sciencedirect.com/science/article/pii/S092765050900070X},
author = {Jere H. Jenkins and Ephraim Fischbach}
}

@misc{mcduffie2020,
      title={Anomalies in Radioactive Decay Rates: A Bibliography of Measurements and Theory}, 
      author={M. H. McDuffie and P. Graham and J. L. Eppele and J. T. Gruenwald and D. Javorsek II and D. E. Krause and E. Fischbach},
      year={2020},
      eprint={2012.00153},
      archivePrefix={arXiv},
      primaryClass={nucl-ex},
      url={https://arxiv.org/abs/2012.00153}, 
}

@article{Emry1972,
author = "Emery, G T",
title = "Perturbation of Nuclear Decay Rates", 
journal= "Annual Review of Nuclear and Particle Science",
year = "1972",
volume = "22",
number = "Volume 22, 1972",
pages = "165-202",
doi = "https://doi.org/10.1146/annurev.ns.22.120172.001121",
url = "https://www.annualreviews.org/content/journals/10.1146/annurev.ns.22.120172.001121",
publisher = "Annual Reviews",
issn = "1545-4134",
type = "Journal Article"
}

@misc{Canberra2024,
title        = {RADIOACTIVITY MEASUREMENT EQUIPMENT},
author       = {MIRION TECHNOLOGIES (CANBERRA) KK},
year         = 2024,
url          = {https://mirionprodstorage.blob.core.windows.net/prod-20220822/cms4_mirion/files/mtkk_a4_brochure_nov2020_ops-978_4X8gwxj.pdf},
howpublished ={\url{https://mirionprodstorage.blob.core.windows.net/prod-20220822/cms4_mirion/files/mtkk_a4_brochure_nov2020_ops-978_4X8gwxj.pdf}},
urldate      = {2024}
}

@misc{Lynx2024,
title        = {Lynx Digital Signal Analyzer},
author       = {MIRION TECHNOLOGIES},
year         = 2024,
url          = {https://mirionprodstorage.blob.core.windows.net/prod-20220822/cms4_mirion/files/pdf/spec-sheets/ops-509_lynx_dsa_spec_rebrand_5.pdf},
howpublished = {\url{https://mirionprodstorage.blob.core.windows.net/prod-20220822/cms4_mirion/files/pdf/spec-sheets/ops-509_lynx_dsa_spec_rebrand_5.pdf}},
urldate      = {2024}
}

@inproceedings{He2015_08,
title={A Study of Background Subtraction Method for NaI(Tl) Instrument Spectrum Based on Adaptive FWHM},
author={Jianfeng He and Hailing Xiao and Yaozong Yang and Jianhui Qu and Hongkun Xu and Liu Lin},
year={2015/08},
booktitle={Proceedings of the 3rd International Conference on Mechanical Engineering and Intelligent Systems (ICMEIS 2015)},
pages={462-468},
issn={2352-5401},
isbn={978-94-62520-98-1},
url={https://doi.org/10.2991/icmeis-15.2015.85},
doi={10.2991/icmeis-15.2015.85},
publisher={Atlantis Press}
}

@article{Szegedi_2021idx,
author = {Szegedi, T. N. and Kiss, G. G. and Mohr, P. and Psaltis, A. and Jacobi, M. and Barnaf\"oldi, G. G. and Sz\"ucs, T. and Gy\"urky, Gy. and Arcones, A.},
title = "{Activation thick target yield measurement of \textsuperscript{100}{Mo} (\ensuremath{\alpha},n) \textsuperscript{103}{Ru} for studying the weak r-process nucleosynthesis}",
doi = "10.1103/PhysRevC.104.035804",
journal = "Phys. Rev. C",
volume = "104",
number = "3",
pages = "035804",
year = "2021"
}

@inproceedings{PoS_Fenyvesi,
title        = {Decay rate measurements with a \textsuperscript{137}{Cs} radioisotope source at {J}ánossy {U}nderground {R}esearch {L}aboratory ({C}sillebérc, {H}ungary)},
author       = {Edit Fenyvesi and Gergely Gábor Barnaföldi and Gábor Gyula Kiss and Dénes Molnár},
year         = 2023,
month        = {},
booktitle    = {Proceedings of Science},
publisher    = {},
address      = {},
series       = {PoS(TAUP2023)},
volume       = {},
number       = 441,
pages        = {347},
editor       = {},
doi = {https://doi.org/10.22323/1.441.0347},
organization = {}
}

@article{BELLOTTI2012114,
title = {Search for time dependence of the \textsuperscript{137}{Cs} decay constant},
journal = {Physics Letters B},
volume = {710},
number = {1},
pages = {114-117},
year = {2012},
issn = {0370-2693},
doi = {https://doi.org/10.1016/j.physletb.2012.02.083},
url = {https://www.sciencedirect.com/science/article/pii/S0370269312002341},
author = {E. Bellotti and C. Broggini and G. {Di Carlo} and M. Laubenstein and R. Menegazzo}
}

@article{BELLOTTI2013116,
title = {Search for correlations between solar flares and decay rate of radioactive nuclei},
journal = {Physics Letters B},
volume = {720},
number = {1},
pages = {116-119},
year = {2013},
issn = {0370-2693},
doi = {https://doi.org/10.1016/j.physletb.2013.02.002},
url = {https://www.sciencedirect.com/science/article/pii/S0370269313001287},
author = {E. Bellotti and C. Broggini and G. {Di Carlo} and M. Laubenstein and R. Menegazzo}
}

@article{MARGINEANU20081501,
title = {The Slanic-Prahova (ROMANIA) underground low-background radiation laboratory},
journal = {Applied Radiation and Isotopes},
volume = {66},
number = {10},
pages = {1501-1506},
year = {2008},
issn = {0969-8043},
doi = {https://doi.org/10.1016/j.apradiso.2008.04.002},
url = {https://www.sciencedirect.com/science/article/pii/S0969804308001814},
author = {R. Margineanu and C. Simion and S. Bercea and O.G. Duliu and D. Gheorghiu and A. Stochioiu and M. Matei}
}

@article{Gyurky2019,
  author    = {Gy{\"u}rky, Gy. and F{\"u}l{\"o}p, Zs. and K{\"a}ppeler, F. and Kiss, G. G. and Wallner, A.},
  title     = {The activation method for cross section measurements in nuclear astrophysics},
  journal   = {The European Physical Journal A},
  year      = {2019},
  volume     = {55},
  number     = {3},
  pages      = {41},
  doi        = {10.1140/epja/i2019-12708-4},
  url        = {https://doi.org/10.1140/epja/i2019-12708-4},
  issn       = {1434-601X}
}

@article{DarkSide,
    author = "Aalseth, C. E. and others",
    collaboration = "DarkSide-20k",
    title = "{DarkSide-20k: A 20 tonne two-phase LAr TPC for direct dark matter detection at LNGS}",
    eprint = "1707.08145",
    archivePrefix = "arXiv",
    primaryClass = "physics.ins-det",
    reportNumber = "FERMILAB-PUB-17-298-PPD",
    doi = "10.1140/epjp/i2018-11973-4",
    journal = "Eur. Phys. J. Plus",
    volume = "133",
    pages = "131",
    year = "2018"
}

@article{CRESST,
  title = {First results from the CRESST-III low-mass dark matter program},
  author = {Abdelhameed, A. H. and Angloher, G. and Bauer, P. and Bento, A. and Bertoldo, E. and Bucci, C. and Canonica, L. and D'Addabbo, A. and Defay, X. and Di Lorenzo, S. and Erb, A. and Feilitzsch, F. v. and Fichtinger, S. and Ferreiro Iachellini, N. and Fuss, A. and Gorla, P. and Hauff, D. and Jochum, J. and Kinast, A. and Kluck, H. and Kraus, H. and Langenk\"amper, A. and Mancuso, M. and Mokina, V. and Mondragon, E. and M\"unster, A. and Olmi, M. and Ortmann, T. and Pagliarone, C. and Pattavina, L. and Petricca, F. and Potzel, W. and Pr\"obst, F. and Reindl, F. and Rothe, J. and Sch\"affner, K. and Schieck, J. and Schipperges, V. and Schmiedmayer, D. and Sch\"onert, S. and Schwertner, C. and Stahlberg, M. and Stodolsky, L. and Strandhagen, C. and Strauss, R. and T\"urko\ifmmode \check{g}\else \v{g}\fi{}lu, C. and Usherov, I. and Willers, M. and Zema, V.},
  collaboration = {CRESST Collaboration},
  journal = {Phys. Rev. D},
  volume = {100},
  issue = {10},
  pages = {102002},
  numpages = {11},
  year = {2019},
  month = {Nov},
  publisher = {American Physical Society},
  doi = {10.1103/PhysRevD.100.102002},
  url = {https://link.aps.org/doi/10.1103/PhysRevD.100.102002}
}

@article{Cano_Ott_2026,
   title={Review of neutron yield from ({$\alpha$},n) reactions: data, methods, and prospects},
   volume={53},
   ISSN={1361-6471},
   url={http://dx.doi.org/10.1088/1361-6471/adeffa},
   DOI={10.1088/1361-6471/adeffa},
   number={2},
   journal={Journal of Physics G: Nuclear and Particle Physics},
   publisher={IOP Publishing},
   author={Cano-Ott, D and Cebrián, S and Dimitriou, P and Gromov, M and Harańczyk, M and Kish, A and Kluck, H and Kudryavtsev, V A and Lazanu, I and Lozza, V and Luzón, G and Mendoza, E and Parvu, M and Pesudo, V and Pocar, A and Santorelli, R and Selvi, M and Westerdale, S and Zuzel, G},
   year={2026},
   month=Feb, pages={023001} }

@article{SNIP1988,
title = {SNIP, a statistics-sensitive background treatment for the quantitative analysis of PIXE spectra in geoscience applications},
journal = {Nuclear Instruments and Methods in Physics Research Section B: Beam Interactions with Materials and Atoms},
volume = {34},
number = {3},
pages = {396-402},
year = {1988},
issn = {0168-583X},
doi = {https://doi.org/10.1016/0168-583X(88)90063-8},
url = {https://www.sciencedirect.com/science/article/pii/0168583X88900638},
author = {C.G. Ryan and E. Clayton and W.L. Griffin and S.H. Sie and D.R. Cousens}
}

@article{Mougeot_2026,
doi = {10.1088/1681-7575/ae3733},
url = {https://doi.org/10.1088/1681-7575/ae3733},
year = {2026},
month = {feb},
publisher = {IOP Publishing},
volume = {63},
number = {1},
pages = {019001},
author = {Mougeot, Xavier and Chechev, Valery P and Dulieu, Christophe and Huang, Xiaolong and Kellett, Mark A and Kuzmenko, Nikolay K and Leblond, Sylvain and Zimmerman, Brian E},
title = {Evaluations of the decay data of 40K, 45Ti, 56Co, 99Tc, 177Lu, 212Pb and 242Am from the Decay Data Evaluation Project (DDEP)-2025},
journal = {Metrologia}
}

@article{Mougeot_2025,
doi = {10.1088/1681-7575/adb9de},
url = {https://dx.doi.org/10.1088/1681-7575/adb9de},
year = {2025},
month = {mar},
publisher = {IOP Publishing},
volume = {62},
number = {2},
pages = {029002},
author = {Mougeot, Xavier and Dulieu, Christophe and Huang, Xiaolong and Kellett, Mark A and Leblond, Sylvain and Wang, Baosong},
title = {Evaluations of the decay data of 137mBa, 137Cs, 151Sm and 225Ac from the Decay Data Evaluation Project (DDEP)-2023},
journal = {Metrologia}
}

@article{Kappeler2011,
  author  = {K{\"a}ppeler, F. and Gallino, R. and Bisterzo, S. and Aoki, W.},
  title   = {The $s$ process: nuclear physics, stellar models, and observations},
  journal = {Reviews of Modern Physics},
  volume  = {83},
  pages   = {157--193},
  year    = {2011},
  doi     = {10.1103/RevModPhys.83.157}
}

@article{Bisterzo2015,
  author  = {Bisterzo, S. and Gallino, R. and K{\"a}ppeler, F. and Wiescher, M. and Imbriani, G. and Straniero, O. and Cristallo, S. and G{\"o}rres, J. and deBoer, R. J.},
  title   = {The branchings of the main $s$-process: their sensitivity to $\alpha$-induced reactions on $^{13}$C and $^{22}$Ne and to the uncertainties of the nuclear network},
  journal = {Monthly Notices of the Royal Astronomical Society},
  volume  = {449},
  pages   = {506--527},
  year    = {2015},
  doi     = {10.1093/mnras/stv271}
}

@article{Mosconi2010,
  author  = {Mosconi, M. and Fujii, K. and Mengoni, A. and others},
  title   = {Neutron physics of the {Re/Os} clock. {I}. {M}easurement of the $(n,\gamma)$ cross sections of $^{186,187,188}${Os} at the {CERN} n\_{TOF} facility},
  journal = {Physical Review C},
  volume  = {82},
  pages   = {015802},
  year    = {2010},
  doi     = {10.1103/PhysRevC.82.015802},
  note    = {n\_TOF Collaboration}
}

@article{Abbondanno2004,
  author  = {Abbondanno, U. and Aerts, G. and Alvarez, H. and others},
  title   = {Neutron capture cross section measurement of $^{151}${Sm} at the {CERN} neutron time of flight facility (n\_{TOF})},
  journal = {Physical Review Letters},
  volume  = {93},
  pages   = {161103},
  year    = {2004},
  doi     = {10.1103/PhysRevLett.93.161103},
  note    = {n\_TOF Collaboration}
}

@article{Reifarth2003,
  author  = {Reifarth, R. and Arlandini, C. and Heil, M. and K{\"a}ppeler, F. and Sedyshev, P. V. and Mengoni, A. and Herman, M. and Rauscher, T. and Gallino, R. and Travaglio, C.},
  title   = {Stellar neutron capture on promethium: implications for the $s$-process neutron density},
  journal = {The Astrophysical Journal},
  volume  = {582},
  pages   = {1251--1262},
  year    = {2003},
  doi     = {10.1086/344718}
}

@article{Pignatari2010,
  author  = {Pignatari, M. and Gallino, R. and Heil, M. and Wiescher, M. and K{\"a}ppeler, F. and Herwig, F. and Bisterzo, S.},
  title   = {The weak $s$-process in massive stars and its dependence on the neutron capture cross sections},
  journal = {The Astrophysical Journal},
  volume  = {710},
  pages   = {1557--1577},
  year    = {2010},
  doi     = {10.1088/0004-637X/710/2/1557}
}

@book{Leo1987,
  author    = {William R. Leo},
  title     = {Techniques for Nuclear and Particle Physics Experiments: A How-to Approach},
  publisher = {Springer-Verlag},
  address   = {Berlin, Heidelberg},
  year      = {1987},
  isbn      = {3-540-17386-2}
}

@article{Tot23,
  title = {Experimental determination of the $^{3}\mathrm{He}(\ensuremath{\alpha},\ensuremath{\gamma})^{7}\mathrm{Be}$ reaction cross section  above the $^{7}\mathrm{Be}$ proton separation threshold},
  author = {T{\'o}th, {\'A}. and Sz{\"u}cs, T. and Szegedi, T. N. and Gy{\"u}rky, Gy. and Hal{\'a}sz, Z. and Kiss, G. G. and F{\"u}l{\"o}p, Zs.},
  journal = {Phys. Rev. C},
  volume = {108},
  issue = {2},
  pages = {025802},
  numpages = {10},
  year = {2023},
  month = {Aug},
  publisher = {American Physical Society},
  doi = {10.1103/PhysRevC.108.025802},
  url = {https://link.aps.org/doi/10.1103/PhysRevC.108.025802}
}

@article{Tot24,
title = {Experimental 7Be production cross section from the Li7(p,n)7Be reaction at Ep = 3.5--13 MeV},
journal = {Nuclear Physics A},
volume = {1041},
pages = {122778},
year = {2024},
issn = {0375-9474},
doi = {https://doi.org/10.1016/j.nuclphysa.2023.122778},
url = {https://www.sciencedirect.com/science/article/pii/S0375947423001811},
author = {{\'A}. T{\'o}th and Gy. Gy{\"u}rky and E. Papp and T. Sz{\"u}cs}
}

\end{document}